\documentclass[
    ,final         
  ]
  {aipproc}

\def\ii{{\,{\rm i}\,}}
\def\dd{{\rm d}}

\newcommand{\Tr}[1]{\:{\rm Tr}\,#1}
\def\e{{\,\rm e}\,}

\def\bea{\begin{eqnarray}}
\def\eea{\end{eqnarray}}
\newcommand{\beq}{\begin{eqnarray}}
\newcommand{\eeq}{\end{eqnarray}}

\layoutstyle{6x9}

\begin{document}

\begin{flushright}
HWM--07--4 \\
EMPG--07--03
\end{flushright}

\title{Symmetries and Renormalization of Noncommutative Field
  Theory
\footnote{Based on plenary lecture delivered at the {\sl
    VI Latin American Symposium on High Energy Physics}, November
  1--8, 2006, Puerto Vallarta, Mexico. To be published in the
  proceedings by American Institute of Physics.}
}


\author{Richard J. Szabo}{
  address={Department of Mathematics and Maxwell Institute for
Mathematical Sciences, Heriot-Watt University, Colin Maclaurin Building,
  Riccarton, Edinburgh EH14 4AS, U.K.}
\footnote{Email: {\tt R.J.Szabo@ma.hw.ac.uk}}
}

\begin{abstract}
 An overview of recent developments in the renormalization and in the
 implementation of spacetime symmetries of noncommutative field theory
 is presented, and argued to be intimately related.
\end{abstract}

\maketitle

\section{Introduction}

\subsection{Noncommutative spaces}

Noncommutative spaces are typically defined in physics by promoting
coordinates $x^i$ of spacetime to hermitean operators $\hat x^i$
satisfying commutation relations of the form $[\hat x^i,\hat
x^j]=\ii\theta^{ij}$, where $\theta^{ij}$ is an antisymmetric tensor
which can be position dependent. Such spaces are believed to be
important for the understanding of quantum gravity. Recent
realizations of noncommutativity in such contexts have been described
for $\kappa$-Minkowski space~\cite{Dimitrijevic:2003wv} and in doubly
special relativity~\cite{Kowalski-Glikman:2002we}, among many others.

A much simpler system in which we can see noncommutative spaces
arising is the Landau problem~\cite{Szabo:2004ic}, which deals with
the motion of charged particles confined to a plane and subjected to a
perpendicularly applied uniform magnetic field $B$. The lagrangian is
$\mathcal{L}_m=\frac m2\,\dot{\bf x}{}^2-e\,\dot{\bf x}\cdot\bf A$,
which in a symmetric gauge for the vector potential $\bf A$ reduces in
the strong field limit $B\gg m$ to
$\mathcal{L}_0=-\frac{e\,B}2\,(\dot x\,y-\dot y\,x)$. This limiting
lagrangian is of first order in time derivatives, implying that upon
canonical quantization the coordinates describe a noncommutative space
$[\hat x,\hat y]=\frac\ii{e\,B}$. This model will play a prominent
role in much of our subsequent discussion.

\subsection{Noncommutative field theory}

The huge interest and large amount of activity came with the seminal
paper~\cite{Seiberg:1999vs} wherein a very natural and precise
physical realization of noncommutative spaces was illustrated, in
complete analogy with the Landau problem above. It was shown that
string theory, in the presence of D-branes and background ``magnetic''
fields, reduces in a particular low-energy limit to field theory on a
noncommutative space~\cite{Douglas:2001ba,Sz1}. A ``toy'' model for this
scenario, and the one this overview will focus on, is the Moyal space
for which the tensor $\theta^{ij}$ is constant and nondegenerate. The
coordinate operators then satisfy Heisenberg commutation relations. In
particular, one has the coordinate uncertainty relations $\Delta
x^i\,\Delta x^j\geq\frac12\,|\theta^{ij}|$. Fields on Moyal spaces are
multiplied using the associative, noncommutative \emph{star-product},
replacing the usual pointwise multiplication $\phi\cdot\psi$ with
\beq
\phi\star\psi:=\phi~\exp\Big(\mbox{$\frac\ii2$}\,
\overleftarrow{\partial}_i\,\theta^{ij}\,\overrightarrow
{\partial}_j\Big)~\psi \ .
\label{starprod}\eeq

Whatever our motivation may be, one of the most interesting and profound
prospects of noncommutative field theory is that it has the
possibility of providing diffeomorphism invariant field theories, and
hence models that we might wish to call noncommutative
gravity~\cite{Szabo:2006wx}. However, in order to provide viable
models of quantum gravity, one is first faced with the technical task
of determining whether or not they make sense. The two main issues are
the precise implementation of spacetime symmetries at both the
classical and quantum level, and the renormalizability of the quantum
field theory. This overview will focus on the very interesting recent
understanding of these two technical issues, and argue that the two
problems are in fact intimately related to one another.

\subsection{Violations of Lorentz invariance}

The constant tensor $\theta^{ij}$ gives a prefered directionality in
space. In string theory, the resulting loss of Lorentz invariance is
due to the expectation value of a background supergravity field. As a
consequence, noncommutative field theory is \emph{not} invariant under
rotations or boosts of localized field configurations within a
\emph{fixed} observer inertial frame of reference, {\sl i.e.} under
particle Lorentz transformations. This observation has been exploited
to construct Lorentz-violating extensions of the standard model
in~\cite{Carroll:2001ws}.

Some possible resolutions to this symmetry breaking are provided by:
\begin{enumerate}
\item Varying $\theta^{ij}$, so that one essentially integrates over
  all possible backgrounds to manifestly reinstate general
  covariance. This has been beautifully implemented
  in~\cite{Doplicher:1994tu} within the framework of $C^*$-algebras,
  but it requires dealing with a host of different noncommutative
  spaces not of the Moyal type which makes extracting
  physical quantities rather difficult.
\item Dimensional reductions of noncommutative gauge theory in higher
  dimensions which induce teleparallel theories of gravity in lower
  dimensions~\cite{Szabo:2006wx,Langmann:2001yr}. However, the
  formalism is somewhat limited in the types of gauge theories of
  gravity one can obtain in this way, and moreover the teleparallelism
  is embedded in a complex way.
\item Exploiting well-known quantum group techniques to
  reinterpret noncommutative field theory as a twist deformed quantum
  field theory~\cite{CKNT1,Wess1}. This realization has been a
  focus of intense activity in recent years and will be the route we
  take in the second part of this overview.
\end{enumerate}

\subsection{UV/IR mixing}

In momentum space, the replacement of the pointwise products of fields
with the star-product (\ref{starprod}) amounts to multiplying the
usual convolution products $\tilde\phi(k)\,\tilde\psi(q)$ of Fourier
transforms by a momentum dependent phase factor $\e^{\ii k\times q}$,
where $k\times q:=\frac12\,k_i\,\theta^{ij}\,q_j$. In a scalar field
theory with polynomial interactions, a typical interaction vertex with
$n$ incoming momenta $k_1,\dots,k_n$ will thus contribute
\beq
\exp\Big(\ii\mbox{$\sum\limits_{I<J}$}\,k_I\times k_J\Big) \ .
\label{intvertex}\eeq
This is effective at energies $E$ with
$E\,\sqrt{\theta}\ll1$. \emph{Planar diagrams} are those Feynman
ribbon graphs which can be drawn without crossing any
lines~\cite{Filk:1996dm,Ishibashi:1999hs}, and their values coincide
with those of the ordinary $\theta=0$ field theory up to possible
phase factors which depend only on external momenta. \emph{Non-planar
  diagrams}, on the other hand, contain
crossings and additional phase factors depending on internal momenta
such as (\ref{intvertex}), whose net result is that an ultraviolet
cutoff $\Lambda$ on the graph induces an effective infrared cutoff
$\Lambda_0=1/\theta\,\Lambda$~\cite{Minwalla:1999px}.

This entangling of momentum scales ruins Wilsonian renormalization,
and two resolutions to this problem are to modify standard
noncommutative field theory to either:
\begin{enumerate}
\item Duality covariant noncommutative field
  theory~\cite{Langmann:2002cc}--\cite{Rivasseau:2005bh}. This is
  described in the next section.
\item Twisted noncommutative field
  theory~\cite{Balachandran:2005pn}. This is explained in the second
  part.
\end{enumerate}

\section{Renormalization}

\subsection{Scalar field theory}

Consider interacting charged scalar fields in four dimensions
with the euclidean action
\beq
S=\int\,\dd^4x~\Big[\phi^\dag\,\big(\partial_i^2+m^2\big)\phi+
\mbox{$\frac\lambda2$}\,\phi^\dag\star\phi\star\phi^\dag\star\phi
\Big]
\label{Sphi4usual}\eeq
where we have used $\int\,\dd^4x~\phi\star\psi=\int\dd^4x~\phi\,\psi$
for any pair of fields, so that the free field theory is unaffected by
the noncommutative deformation. The quantum field theory is given by
the Green's functions which are defined as the $2n$-point correlation
functions
\beq
G_n(x_I,y_I)=\big\langle\phi^\dag(x_1)\,\phi(y_1)\cdots
\phi^\dag(x_n)\,\phi(y_n)\big\rangle \ .
\label{GnxIyI}\eeq
As an example, consider the contribution of the one-loop tadpole graph
to the two-point function in the case of real scalar fields
$\phi=\phi^\dag$.\footnote{The case of complex scalar fields is much
  more subtle, as shown in~\cite{Aref'eva:2000hq}.} If the external
legs carry momentum $p$, then the planar tadpole diagram gives the
momentum space Feynman integral
\beq
\frac\lambda6\,\int\,\frac{\dd^4k}{(2\pi)^4}~\frac1{k^2+m^2}
\label{planartad}\eeq
which exhibits the standard ultraviolet divergence of scalar
$\phi^4$-theory in four dimensions. On the other hand, the non-planar
tadpole diagram gives
\beq
\frac\lambda{12}\,\int\,\frac{\dd^4k}{(2\pi)^4}~\frac
{\e^{2\ii k\times p}}{k^2+m^2}=\frac\lambda{48\pi^2}\,
\sqrt{\frac{m^2}{(\theta\,p)^2}}~K_1\Big(\sqrt{m^2\,(
\theta\,p)^2}~\Big)
\label{nonplanartad}\eeq
which diverges as $(\theta\,p)^{-2}$ as $p\to0$. The model is thus
\emph{not} renormalizable, because the quantum field theory is not
covariant under ``UV/IR duality''~\cite{Grosse:2004yu} as we now
explain.

\subsection{UV/IR duality}

To resolve the problems associated with UV/IR mixing, we introduce a
covariant version of noncommutative field theory in which the
ultraviolet and infrared regimes are
indistinguishable~\cite{Langmann:2002cc}. It is defined by modifying
the action (\ref{Sphi4usual}) to
\beq
S=\int\,\dd^4x~\sqrt{g}~\Big[\phi^\dag\,\big(g^{ij}\,D_i\,D_j+m^2
\big)\phi+\mbox{$\frac\lambda2$}\,\big(\phi^\dag\star\phi\big)^2\Big]
\ ,
\label{dualitySphi4}\eeq
where $g_{ij}$ is a constant metric tensor and
$D_i=\partial_i-\frac\ii2\,B_{ij}\,x^j$ are gauge
covariant derivatives in a magnetic background characterized by
another nondegenerate, constant antisymmetric tensor $B_{ij}$. The new
derivatives obey the commutation relations $[D_i,D_j]=-2\ii B_{ij}$,
so that the modification of the kinetic term in (\ref{dualitySphi4})
can be thought of as inducing a ``noncommutative momentum space''. The
quantum field theory now has a duality under Fourier transformation of
the fields, given by the covariant transformation rules
\bea
S[\phi]\big|_{\lambda,g,B,\theta}&=&|\det B|\,S[\,\tilde\phi\,]
\big|_{\tilde\lambda,\tilde g,\tilde B,\tilde\theta} \ , \nonumber\\
\tilde G_n(\theta^{-1}\,x_I,\theta^{-1}\,y_I)
\big|_{\lambda,g,B,\theta}&=&|\det B|^{n/2}\,G_n(x_I,y_I)
\big|_{\tilde\lambda,\tilde g,\tilde B,\tilde\theta}
\label{UVIRduality}\eeq
where the dual parameters are defined by
$\tilde\lambda=\lambda/\sqrt{\det\theta}$, $\tilde
g=-B^{-1}\,g\,B^{-1}$, $\tilde B=B^{-1}$ and
$\tilde\theta=\theta^{-1}$.

\subsection{Renormalization of the duality-covariant field theory}

The key to the renormalization of the duality-covariant model is the
use of a ``matrix basis'' for the expansion of fields given
by~\cite{Grosse:2004yu,Langmann:2003if}
\beq
\phi(x)=\sum_{\ell,n\in{\bf N}_0^2}\,\phi_{\ell n}^\dag~
\varphi_{\ell n}(x) \ ,
\label{matrixbasis}\eeq
where $\varphi_{\ell n}$ are the ``Landau wavefunctions'' which at
$B=\theta^{-1}$ are the eigenfunctions of the kinetic operator in
(\ref{dualitySphi4}) with $-D_i^2\varphi_{\ell n}={\rm
  Pf}(\theta)^{-1}\,(\ell_1+\ell_2-1)\,\varphi_{\ell n}$. They form a
complete orthornormal system of $L^2$-fields which multiply together
like matrix units $|\ell_1,\ell_2\rangle\langle n_1,n_2|$ under the
star-product as $\varphi_{\ell n}\star\varphi_{\ell'n'}={\rm
  Pf}(4\pi\,\theta)^{-1}\,\delta_{n\ell'}~\varphi_{\ell n'}$. The
noncommutative field theory (\ref{dualitySphi4}) then becomes an
infinite-dimensional complex matrix model
\beq
S=\Tr\Big(\phi^\dag\,{\cal G}\phi+\mbox{$\frac\lambda2$}~
{\rm Pf}(4\pi\,\theta)\,\big(\phi^\dag\,\phi\big)^2\Big) \ .
\label{matrixmodel}\eeq
One of the beautiful calculations performed in~\cite{Grosse:2004yu} is
that of the propagator ${\cal G}^{-1}$ as the formal inverse of the
infinite matrix ${\cal G}=({\cal G}_{\ell n})$, which uses
hypergeometric Meixner $q$-polynomials. The natural ultraviolet cutoff
is now on the matrix dimension as $\ell_1+\ell_2,n_1+n_2\leq N$. One
then slices the propagator in the renormalization group with sharp
bounds on the matrix indices $\ell,n$. By using the Wilson-Polchinski
renormalization group equations, one proves in this way that the
duality-covariant field theory is renormalizable to all
orders~\cite{Grosse:2004yu}.

\subsection{Beyond perturbation theory}

The beta-functions of the couplings $B$ and $\lambda$ in the
duality-covariant field theory have been computed to all orders
in~\cite{Grosse:2004by,Disertori:2006nq}. They are of the
usual sign for any magnetic background $B\neq\theta^{-1}$. At the
special point $B=\theta^{-1}$ a number of remarkable things happen:
\begin{itemize}
\item $\beta_B=\beta_\lambda=0$, hence the renormalized coupling
  flows to a finite bare coupling and the field theory is
  asymptotically safe. This allows in principle a nonperturbative
  construction of the quantum field theory.
\item The noncommutative quantum field theory is completely duality
  \emph{invariant}.
\item The matrix $\cal G$ in (\ref{matrixmodel}) is diagonal with
  entries the Landau energies, and the field theory becomes an exactly
  solvable large $N$ matrix model with a huge unitary symmetry
  $\phi\mapsto U\,\phi\,U^{-1}$, $U\in U(N)$ reflecting the degeneracy
  of Landau levels. In terms of fields this symmetry corresponds to
  the transformations $\phi(x)\mapsto(U\star\phi\star U^\dag)(x)$ with
  $U\star U^\dag=U^\dag\star U=1$, which generate the infinite unitary
  group $U(\infty)$ representing ``deformed'' canonical
  transformations of the spacetime~\cite{Lizzi:2001nd}.
\end{itemize}
Some of these considerations have been generalized to noncommutative
$\phi^3$-theory by mapping it onto the Kontsevich matrix
model~\cite{Grosse:2006qv}, and to the noncommutative Gross-Neveu
model~\cite{Vignes-Tourneret:2006nb}. However, there are difficulties
associated with the \emph{nonperturbative} renormalizability of the
self-dual model~\cite{Langmann:2003if}, due to the overly large
$U(\infty)$ symmetry which appears to kill all non-trivial
dynamics,\footnote{Of course in the real case there is no $U(\infty)$
  symmetry which kills the dynamics. In~\cite{Bietenholz:2004xs}
  numerical evidence is provided for the renormalizability of a
  somewhat different class of noncommutative scalar field theories.}
and we must search for an alternative way to implement the
symplectomorphism symmetry.

\section{Twisted Symmetries}

\subsection{Twist deformations}

Suppose that $X$ is an infinitesimal symmetry transformation of
fields, denoted $\phi\mapsto X\triangleright\phi$. Then the action of
$X$ on tensor products of fields is implemented by a \emph{coproduct}
$\Delta$ with
$\phi\otimes\psi\mapsto\Delta(X)\triangleright(\phi\otimes\psi)$. The
\emph{primitive} coproduct $\Delta=\Delta_0$, with
$\Delta_0(X)=X\otimes1+1\otimes X$, is covariant with respect to the
pointwise product $m_0(\phi\otimes\psi)=\phi\cdot\psi$ in the sense
that
\beq
X\triangleright
m_0(\phi\otimes\psi)=m_0\,\Delta_0(X)\triangleright
(\phi\otimes\psi) \ .
\label{covprimitive}\eeq
For example, for the translation generator $X=P_i=-\ii\partial_i$ the
covariance (\ref{covprimitive}) is just the usual Leibniz rule
$P_i(\phi\cdot\psi)=(P_i\phi)\cdot\psi+\phi\cdot(P_i\psi)$. While
$\Delta_0$ is \emph{not} covariant with respect to the star-product,
the \emph{twist deformed} coproduct $\Delta=\Delta_\theta$
\emph{is}~\cite{CKNT1,Wess1},\cite{Drinfeld:1989st}--\cite{Aschieri:2005yw}.
It is defined by rewriting the star-product (\ref{starprod}) as
$\phi\star\psi=m_\theta(\phi\otimes\psi):=m_0({\cal
  F}\phi\otimes\psi)$, where
\beq
{\cal F}=\exp\Big(-\mbox{$\frac\ii2$}\,\theta^{ij}\,P_i\otimes P_j\Big)
\label{twist}\eeq
is an abelian Drinfeld twist, and forming the similarity
transformation
\beq
\Delta_\theta(X)={\cal F}^{-1}\,\Delta_0(X)\,{\cal F}=
{\cal F}^{-1}\,(X\otimes1+1\otimes X){\cal F} \ .
\label{twistcoprod}\eeq

\subsection{Twisted spacetime symmetries}

Using (\ref{twistcoprod}) one can straightforwardly work out twisted
Poincar\'e transformations generated by the usual linear and angular
momentum operators $P_i=-\ii\partial_i$ and
$M_{ij}=-\ii(x_i\,\partial_j-x_j\,\partial_i)$. One finds
$\Delta_\theta(P_i)=\Delta_0(P_i)$, reflecting the fact that
noncommutative field theory is translationally invariant, whereas
\beq
\Delta_\theta(M_{ij})=\Delta_0(M_{ij})+\mbox{$\frac\ii2$}\,
\theta^{kl}\,(\eta_{ik}\,P_j-\eta_{jk}\,P_i)\otimes P_l+
\mbox{$\frac\ii2$}\,\theta^{kl}\,P_k\otimes(
\eta_{il}\,P_j-\eta_{jl}\,P_i) \ ,
\label{DeltaMij}\eeq
reflecting that it is not invariant under boosts or
rotations. However, from (\ref{DeltaMij}) one finds that
noncommutative field theory is invariant under {\it twisted} particle
transformations, because
\beq
M_{kl}\triangleright[x^i,x^j]_\star=m_\theta\,\Delta_\theta(M_{kl})
\triangleright(x^i\otimes x^j-x^j\otimes x^i)=0
\label{twistedpart}\eeq
is equivalent to $M_{kl}\triangleright\theta^{ij}=0$. This symmetry
can be generalized to linear affine transformations
$x\mapsto L\,x+a$, $\theta\mapsto L\,\theta\,L^\top$ using covariance
of the Moyal star-product~\cite{Gracia-Bondia:2006yj}. One can in fact
generalize the symmetry to twisted diffeomorphisms, generated by
arbitrary smooth vector fields
$X=X^i(x)~\partial_i$~\cite{Aschieri:2005yw}. One generically has
$\Delta_\theta(X)\neq\Delta_0(X)$, but one can construct a ``twisted''
tensor calculus such that star-products of tensor fields transform as
tensors. However, only unimodular $U(\infty)$ twisted transformations,
with $\partial_iX^i=0$, preserve action functionals of noncommutative
field theories~\cite{RS1}.

\subsection{Twisted noncommutative quantum field theory}

Given a set of one-particle wavefunctions $\phi(x)$, two-particle
wavefunctions are constructed as $(\phi\otimes\psi)(x_1,x_2)$. The flip map
$\sigma_0(\phi\otimes\psi)=\psi\otimes\phi$ is only superselected when
$\theta=0$, as then $\sigma_0\,\Delta_0=\Delta_0\,\sigma_0$. For
$\theta\neq0$, we use instead the ``twisted'' flip operator
$\sigma_\theta={\cal F}^{-1}\,\sigma_0\,{\cal F}={\cal
  F}^{-2}\,\sigma_0$ with $\sigma_\theta^2=1\otimes1$ and
$\sigma_\theta\,\Delta_\theta=\Delta_\theta\,\sigma_\theta$ (${\cal
  F}^{-2}$ is the corresponding R-matrix). If $c(p)$ are the usual
(bosonic or fermionic) oscillators of the undeformed quantum field
theory, then this modifies them to the twisted oscillators
\beq
a(p)=c(p)~\exp\Big(\mbox{$\frac\ii2$}\,\theta^{ij}\,p_i~P_j\Big)
\label{twistoscs}\eeq
with the commutation relations $a(p)\,a(q)=\pm~\e^{2\ii p\times
  q}\,a(q)\,a(p)$. In particular, on free quantum fields the creation
parts obey
\beq
\phi^{(+)}(x_1)\,\phi^{(+)}(x_2)=\pm~\exp\Big(\ii\theta^{ij}\,
\mbox{$\frac\partial{\partial x_2^i}\,\frac\partial{\partial x_1^j}$}
\Big)~\phi^{(+)}(x_2)\,\phi^{(+)}(x_1) \ .
\label{freequantumcomms}\eeq
This modifies the ordinary Feynman path integral measure
$\prod_x\,\dd\phi(x)$ and defines a \emph{braided quantum field
  theory} with covariant Wick expansions~\cite{Oeckl:1999zu}. The two
main consequences of these arguments are:
\begin{enumerate}
\item For a spin~$0$ field with interaction hamiltonian density of the
  form ${\cal H}_{\rm I}(x)=\phi(x)\star\cdots\star\phi(x)$, the
  corresponding S-matrix is independent of $\theta$ and hence there is
  \emph{no} UV/IR mixing in twisted quantum field
  theory~\cite{Balachandran:2005pn,Oeckl:2000eg}.
\item Even \emph{free} twisted noncommutative field theory is
  \emph{non-local}, because one has $\langle
  q|[\phi(x),\phi(y)]|p\rangle\neq0$ for $q\neq p$ and space-like
  separations~\cite{Balachandran:2006pi}.
\end{enumerate}

\subsection{Twisted Noether symmetries}

Twisted diffeomorphisms do not appear to be \emph{bonafide} physical
symmetries, because they do not act solely on fields. They modify the
usual Leibniz rule (represented by the primitive coproduct $\Delta_0$)
through transformation of the star-product as
\beq
\delta_X(\phi\star\psi):=m_\theta\,\Delta_\theta(X)\triangleright
(\phi\otimes\psi)=(\delta_X\phi)\star\psi+\phi
\star(\delta_X\psi)+\phi(\delta_X\star)\psi \ .
\label{starprodtransf}\eeq
The extra variation seems to present an obstacle to application of the
\emph{standard} Noether procedure and of the Ward identities in the
quantum field theory. One resolution, proposed
in~\cite{Agostini:2006nc} (see also~\cite{RS1}), is to use a proper
covariant noncommutative differential calculus to perform the Noether
analysis relating fields and conserved charges. However, the physical
origins of these symmetries are unclear, and in particular string
theory is unable to account for twisted
diffeomorphisms~\cite{Alvarez-Gaume:2006bn}. The brane-induced
low-energy dynamics of closed string theory in the presence of a
$B$-field is \emph{much} richer than any noncommutative theory of
gravity based solely on star-products.

\section{Outlook}

We have described how two notorious problems of noncommutative field
theory can be resolved by at least two rather drastic modifications of the
underlying models, leading to the covariant and twisted noncommutative
field theories. Both models are free from UV/IR mixing, possess a
$U(\infty)$ symplectomorphism symmetry, and have underlying non-local
free field theories. The main outstanding problems are to find natural
\emph{physical} origins for covariant and twisted noncommutative field
theories, and to resolve the ambiguities in the definitions of scaling
limits and of correlation functions. Another open problem is to
generalize these modifications to the framework of gauge theories,
wherein UV/IR mixing is logarithmic and is associated to open Wilson
line operators coupling gravity to D0-branes~\cite{Sz1}.

\begin{theacknowledgments}

This work was supported in part by the EU-RTN Network Grant
MRTN-CT-2004-005104. The author gratefully thanks David~Vergara and
the other organisors of the Silafae conference for the hospitality in
a beautiful and stimulating atmosphere.

\end{theacknowledgments}

\end{document}